\documentclass[aps,prl,showpacs,showkeys,twocolumn,superscriptaddress]{revtex4}

\usepackage[dvips]{graphicx}

\begin{document}
\bibliographystyle{try}
\topmargin 0.1cm
\title{{\large\bf
Cascade production in the reactions $\gamma p \rightarrow K^+ K^+ (X)$ and $\gamma p \rightarrow K^+ K^+ \pi^- (X)$ }}

\newcommand*{\JLAB }{ Thomas Jefferson National Accelerator Facility, Newport News, Virginia 23606} 
\affiliation{\JLAB } 
\newcommand*{\INFNGE }{ INFN, Sezione di Genova, 16146 Genova, Italy} 
\affiliation{\INFNGE } 
\newcommand*{\RPI }{ Rensselaer Polytechnic Institute, Troy, New York 12180-3590} 
\affiliation{\RPI } 
\newcommand*{\ANL}{Argonne National Laboratory}
\affiliation{\ANL}
\newcommand*{\ASU}{Arizona State University, Tempe, Arizona 85287-1504}
\affiliation{\ASU}
\newcommand*{\UCLA}{University of California at Los Angeles, Los Angeles, California  90095-1547}
\affiliation{\UCLA}
\newcommand*{\CSU}{California State University, Dominguez Hills, Carson, CA 90747}
\affiliation{\CSU}
\newcommand*{\CMU}{Carnegie Mellon University, Pittsburgh, Pennsylvania 15213}
\affiliation{\CMU}
\newcommand*{\CUA}{Catholic University of America, Washington, D.C. 20064}
\affiliation{\CUA}
\newcommand*{\SACLAY}{CEA-Saclay, Service de Physique Nucl\'eaire, 91191 Gif-sur-Yvette, France}
\affiliation{\SACLAY}
\newcommand*{\CNU}{Christopher Newport University, Newport News, Virginia 23606}
\affiliation{\CNU}
\newcommand*{\UCONN}{University of Connecticut, Storrs, Connecticut 06269}
\affiliation{\UCONN}
\newcommand*{\ECOSSEE}{Edinburgh University, Edinburgh EH9 3JZ, United Kingdom}
\affiliation{\ECOSSEE}
\newcommand*{\FU}{Fairfield University, Fairfield CT 06824}
\affiliation{\FU}
\newcommand*{\FIU}{Florida International University, Miami, Florida 33199}
\affiliation{\FIU}
\newcommand*{\FSU}{Florida State University, Tallahassee, Florida 32306}
\affiliation{\FSU}
\newcommand*{\GWU}{The George Washington University, Washington, DC 20052}
\affiliation{\GWU}
\newcommand*{\ECOSSEG}{University of Glasgow, Glasgow G12 8QQ, United Kingdom}
\affiliation{\ECOSSEG}
\newcommand*{\ISU}{Idaho State University, Pocatello, Idaho 83209}
\affiliation{\ISU}
\newcommand*{\INFNFR}{INFN, Laboratori Nazionali di Frascati, 00044 Frascati, Italy}
\affiliation{\INFNFR}
\newcommand*{\ORSAY}{Institut de Physique Nucleaire ORSAY, Orsay, France}
\affiliation{\ORSAY}
\newcommand*{\ITEP}{Institute of Theoretical and Experimental Physics, Moscow, 117259, Russia}
\affiliation{\ITEP}
\newcommand*{\JMU}{James Madison University, Harrisonburg, Virginia 22807}
\affiliation{\JMU}
\newcommand*{\KYUNGPOOK}{Kyungpook National University, Daegu 702-701, South Korea}
\affiliation{\KYUNGPOOK}
\newcommand*{\UMASS}{University of Massachusetts, Amherst, Massachusetts  01003}
\affiliation{\UMASS}
\newcommand*{\MOSCOW}{Moscow State University, Skobeltsyn Nuclear Physics Institute, 119899 Moscow, Russia}
\affiliation{\MOSCOW}
\newcommand*{\UNH}{University of New Hampshire, Durham, New Hampshire 03824-3568}
\affiliation{\UNH}
\newcommand*{\NSU}{Norfolk State University, Norfolk, Virginia 23504}
\affiliation{\NSU}
\newcommand*{\OHIOU}{Ohio University, Athens, Ohio  45701}
\affiliation{\OHIOU}
\newcommand*{\ODU}{Old Dominion University, Norfolk, Virginia 23529}
\affiliation{\ODU}
\newcommand*{\RICE}{Rice University, Houston, Texas 77005-1892}
\affiliation{\RICE}
\newcommand*{\URICH}{University of Richmond, Richmond, Virginia 23173}
\affiliation{\URICH}
\newcommand*{\SCAROLINA}{University of South Carolina, Columbia, South Carolina 29208}
\affiliation{\SCAROLINA}
\newcommand*{\TRIUMF}{TRIUMF, VanCouver, BC V6T 2A3, Canada}
\affiliation{\TRIUMF}
\newcommand*{\UNIONC}{Union College, Schenectady, NY 12308}
\affiliation{\UNIONC}
\newcommand*{\VT}{Virginia Polytechnic Institute and State University, Blacksburg, Virginia   24061-0435}
\affiliation{\VT}
\newcommand*{\VIRGINIA}{University of Virginia, Charlottesville, Virginia 22901}
\affiliation{\VIRGINIA}
\newcommand*{\WM}{College of William and Mary, Williamsburg, Virginia 23187-8795}
\affiliation{\WM}
\newcommand*{\YEREVAN}{Yerevan Physics Institute, 375036 Yerevan, Armenia}
\affiliation{\YEREVAN}
\newcommand*{\CORRES}{lguo@jlab.org; Corresponding author.}
\newcommand*{\NOWUNH}{University of New Hampshire, Durham, New Hampshire 03824-3568}
\newcommand*{\NOWUMASS}{University of Massachusetts, Amherst, Massachusetts  01003}
\newcommand*{\NOWMIT}{Massachusetts Institute of Technology, Cambridge, Massachusetts  02139-4307}
\newcommand*{\NOWCUA}{Catholic University of America, Washington, D.C. 20064}
\newcommand*{\NOWECOSSEE}{Edinburgh University, Edinburgh EH9 3JZ, United Kingdom}
\newcommand*{\NOWGEISSEN}{Physikalisches Institut der Universitaet Giessen, 35392 Giessen, Germany}

\author{L.~Guo}
\altaffiliation[Electronic address: ]{\CORRES}
\affiliation{\JLAB}
\author{D.P.~Weygand}
\affiliation{\JLAB}
\author {M.~Battaglieri} 
\affiliation{\INFNGE}
\author {R.~De~Vita} 
\affiliation{\INFNGE}
\author{V.~Kubarovsky}
\affiliation{\RPI }
\author{P.~Stoler}
\affiliation{\RPI }
\author {M.J.~Amaryan} 
\affiliation{\ODU}
\author {P.~Ambrozewicz} 
\affiliation{\FIU}
\author {M.~Anghinolfi} 
\affiliation{\INFNGE}
\author {G.~Asryan} 
\affiliation{\YEREVAN}
% ...

\collaboration{The CLAS Collaboration}
     \noaffiliation
\date{\today}


\begin{abstract}                

Photoproduction of the cascade resonances has been investigated in the reactions $\gamma p \rightarrow K^+ K^+ (X)$ and  $\gamma p \rightarrow K^+ K^+ \pi^- (X)$. The mass splitting of the ground state ($\Xi^-, \Xi^0$) doublet is measured to be $5.4\pm 1.8$~MeV/{\it c}$^2$, consistent with existing measurements. The differential (total) cross sections for the $\Xi^{-}$ have been determined for photon beam energies from 2.75 to 3.85 (4.75)~GeV, and are consistent with a production mechanism of $Y^*\rightarrow K^+\Xi^-$ through a $t$-channel process. The reaction $\gamma p \rightarrow K^+ K^+ \pi^-[\Xi^0]$ has also been investigated in search of excited cascade resonances. No significant signal of excited cascade states other than the $\Xi^-(1530)$ is observed. The cross section results of the $\Xi^-(1530)$ have also been obtained for photon beam energies from 3.35 to 4.75~GeV. 

\end{abstract}
\maketitle
\newpage

\vspace{-5mm}
\section{Introduction}
\vspace{-3mm}

Hadron spectroscopy is an essential experimental means of accessing fundamental parameters of QCD such as quark masses. The average of the baryon ground state isospin multiplet ($N, \Sigma, \Delta, \Xi, \Sigma_c, \Xi_c$) mass differences yields a value of $m_d - m_u = +(2.8\pm 0.3)$~MeV/{\it c}$^2$~\cite{NEFKENS}, with the $\Xi$ ground state doublet being the most intriguing. 

The current global measurement of the mass difference between the $\Xi^0(uss)$ and $\Xi^-(dss)$ is $6.48\pm0.24$~MeV/{\it c}$^2$ according to the PDG~\cite{YAO}, considerably larger than that of the other multiplets. A calculation on the QCD lattice~\cite{Duncan} gives a result of $5.68\pm0.24$~MeV/{\it c}$^2$, while a calculation  based on radiative corrections to the quark model~\cite{Delbourgo} gives 6.10~MeV/{\it c}$^2$. Experimentally, however, only one measurement of the $\Xi^0$ mass has more than 50 events~\cite{Fanti}.

Compared with non-strange baryons and $S=-1$ hyperon states, the $\Xi$ resonances are generally under-explored. Only two ground state cascades, the octet member $\Xi$ and the decuplet member $\Xi(1530)$, have four-star status in the PDG~\cite{YAO}, with four other three-star candidates. The lack of data is mainly due to smaller $\Xi^{(*)}$ cross sections than the  $S=0$ and $-1$ baryons, and cascade resonances cannot be produced through direct formation.
 More than 20 $N^*$ and $\Delta^*$ resonances are rated with at least three stars in the PDG~\cite{YAO}.  Flavor $SU(3)$ symmetry predicts as many $\Xi$ resonances as $N^*$ and $\Delta^*$ states combined, suggesting that  many more cascade resonances remain undiscovered. Of the six $\Xi$ states that have at least three-star ratings in the PDG, only three have spin-parity ($J^P$) determined: $\Xi(1320)\frac{1}{2}^+$, $\Xi(1530)\frac{3}{2}^+$, $\Xi(1820)\frac{3}{2}^-$. 

\begin{figure}[htbp]
\vspace{56mm}
{\includegraphics{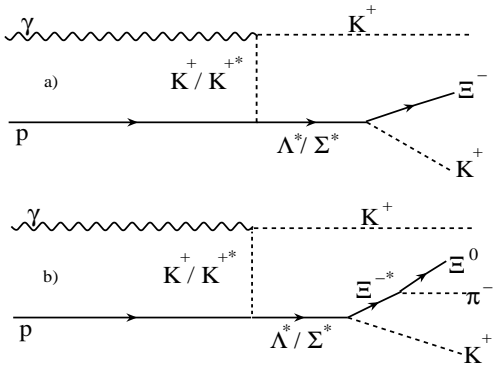}}
\vspace{2mm}
\caption{Possible photoproduction mechanisms of $\Xi$ ground states through intermediate hyperon resonances produced in a $t$-channel process. a)$\Xi^-$ production; b) $\Xi^0$ production.}
\label{dia}
\end{figure}

In general, the production mechanisms of the cascade resonances remain unclear. Kaon and hyperon beam experiments conducted to investigate cascade spectroscopy suffer from either low intensity or high combinatorial background. Results from earlier kaon beam experiments indicate that it is possible to produce the $\Xi$ ground state through the decay of high-mass $Y^*$ states~\cite{Muller, Burgun, Tripp, Litchfield}. It is therefore possible to produce cascade resonances through $t$-channel photoproduction of hyperon resonances as indicated in Fig.~\ref{dia}.

By using tagged photons incident on a proton target, it has been demonstrated that cascade production can be investigated through exclusive reactions, such as $\gamma p \rightarrow K^+ K^+ (X)$~\cite{Price} in CLAS.  Prior to this publication, only two groups have reported measurements of cascade photoproduction, both in the inclusive reaction $\gamma p \rightarrow \Xi^- X$ by reconstructing the $\Xi^-$ from the  decay  $\Xi^{-}\rightarrow\Lambda\pi^-\rightarrow p\pi^-\pi^-$. The CERN SPS experiment with the Omega spectrometer~\cite{Aston} measured a cross section of $28\pm9$~nb for the kinematical range $x_F( = 2p^*_{\parallel}/\sqrt{s})>-0.3$, using a tagged photon beam in the energy range $20-70$~GeV. However, the SLAC 1-m hydrogen bubble chamber experiment~\cite{Abe} using a 20~GeV photon beam reported a much higher cross section of $94\pm13$~nb in the same $x_F$ range, with a total cross section of $117\pm17$~nb. 

The SLAC results showed that the $x_F$ distribution of the $\Xi^-$ events peaks around $-\frac{1}{3}$, consistent with a quark-diquark fusion production mechanism~\cite{DONNACHIE}, in which the cascade has one out of three quarks in common with the proton. However, such a model is more appropriate for inclusive reactions at high energies where partonic degrees of freedom are more relevant, and it is not applicable for exclusive reactions at low to intermediate energies compared with the threshold ($E_{\gamma}^{thres}=2.37$~GeV). Recently, Nakayama {\it et~al.}~\cite{Oh} developed a $\Xi$ production model for the exclusive reaction $\gamma N \rightarrow K K \Xi$ from an effective Lagrangian that incorporates various $t$, $u$, and $s$-channel processes, taking into account intermediate hyperon and nucleon resonances (details of the model will be discussed later in this paper). The validity of the model should be checked by comparing its predictions of the model with experimental data.

In this paper, the mass difference of the $\Xi$ doublet, the cross sections of the $\Xi^-$ and $\Xi^-(1530)$ are reported and compared with Ref.~\cite{Oh}. The possibility of producing other excited cascade states in photon-proton reaction is also discussed.

\vspace{-5mm}
\section{Experiment}
\vspace{-3mm}
A new, large statistics data set, with an integrated luminosity of 70~pb$^{-1}$, was collected at CLAS~\cite{CLAS} from May to July 2004 using a tagged photon beam~\cite{TAGGER} incident on a proton target. This data set is mostly in the energy range of 1.6-3.85~GeV with the primary electron beam energy ($E_0$) of 4~GeV. About $5\%$ of the data were collected with $E_0=5$~GeV. The target consists of a 40-cm-long cylindrical cell containing liquid hydrogen. Momentum information for charged particles was obtained via tracking through three regions of multi-wire drift chambers~\cite{DC} inside a toroidal magnetic field ($\sim 0.5$ T), generated by six superconducting coils. Time-of-flight (TOF) scintillators were used for charged hadron identification~\cite{TOF}. The interaction time between the incoming photon and the target was measured by the Start Counter ~\cite{ST}, consisting of 24 strips of 2.2~mm thick plastic scintillators surrounding the target cell. Coincidences between the photon tagger and two charged particles in the CLAS detector triggered the events. 

\begin{figure}[htbp]
\vspace{56mm}
{\includegraphics{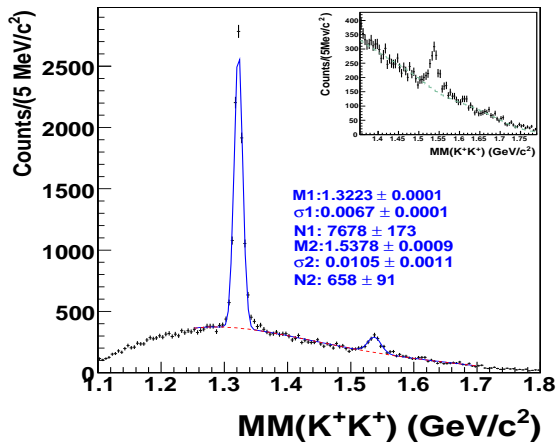}}
\vspace{2mm}
\caption{$MM(K^+K^+)$ distribution for $E_{\gamma} > 2.6$~GeV fitted with two Gaussian functions and an empirical background shape with adjustable normalization (M: mean of the Gaussian peak position, $\sigma$: width of the Gaussian signal, N: number of events in the peak); Inset:$MM(K^+K^+)$ distribution enlarged for the 1.36-1.79~GeV/{\it c}$^2$ region, the dashed lines show the empirical background shape from $K^-$ events normalized to the region of 1.36-1.5~GeV/{\it c}$^2$.}
\label{mmkk}
\end{figure}

Cascade states can be identified via missing mass, through the reaction $\gamma p \rightarrow K^+ K^+ (X)$, or via the decay $\Xi^{*-}\rightarrow\Xi^0\pi^-$ through the reaction $\gamma p \rightarrow K^+ K^+ \pi^- (X)$. In the reaction $\gamma p \rightarrow K^+ K^+ (X)$, the double-strangeness is tagged by the two positive kaons detected by CLAS, and the cascade resonances are observed in the $K^+K^+$ missing mass spectrum (Fig.~\ref{mmkk}). Without the more stringent particle identification criteria that were applied in Fig.~\ref{mmkk}, ({\it i.e}, the kaon vertex time determined by the TOF is within 1~ns of the photon time given by the RF), more than 12000 $\Xi^-$ were observed~\cite{Guo}. After the tighter detector timing cut was applied, about 7700 $\Xi^-$ events are identified for the photon energy range of 2.6 to 4.75~GeV. There is no $\Xi^-$ signal for $E_{\gamma}<$~2.6~GeV, most likely due to low acceptance.

The $\Xi^-(1530)$ is clearly present in the spectrum, with about $700$ events (Fig.~\ref{mmkk}). Events with an additional $K^-$ detected are used as an empirical background, since the background is dominated by reactions such as $\gamma p \rightarrow K^+ K^- p$ or $\gamma p \rightarrow K^+ K^- \pi^+ n$, with the proton or $\pi^+$ misidentified as a $K^+$ (potential background processes such as $\gamma p\rightarrow \phi \Lambda K^+$ were explored and found to be insignificant). The background is then smoothed and normalized to the region between the $\Xi^-$ and the $\Xi^-(1530)$ resonances ($1.36-1.5$~GeV/{\it c}$^2$) in the $MM(K^+K^+)$ distribution (Fig.~\ref{mmkk}, inset). The $\Xi^-$ mass is determined to be $1322.3\pm 0.1 \pm 1.2$~MeV/{\it c}$^2$, slightly higher than the PDG~\cite{YAO} value but within errors. The systematic uncertainty is derived from studying the variation of the fitted mass centroid as a function of $E_{\gamma}$. The $\Xi^-$ width is $6.7\pm 0.1$~MeV/{\it c}$^2$, and is consistent with the missing mass resolution of CLAS as expected from simulation. It is mostly dependent on the resolution of the photon energy measurement, which is typically around 0.1$\%$ of the incident photon energy~\cite{TAGE}.

\vspace{-5mm}
\section{$\Xi^-$ cross section results}
\vspace{-3mm}
The observed $\Xi^-$ events in this work represent the highest statistics seen in exclusive photoproduction to date. It is possible to probe the production mechanism through various differential cross sections, such as $d\sigma/dM(K^+\Xi^-)$, $d\sigma/dM(K^+K^+)$, $d\sigma/d\cos\theta^*_{\Xi^-}$, and $d\sigma/d\cos\theta^*_{K^+}$. To extract the cross section for the $\Xi^-$, a detailed simulation has been carried out. Assuming a $t$-channel process, the reaction $\gamma p \rightarrow K^+ Y^*, Y^*\rightarrow K^+ \Xi^-$ was simulated. Although earlier experiments have reported the possible observation of $Y^*\rightarrow \Xi K$ for the $\Sigma(2030)(J^P=\frac{7}{2}^+)$ and $\Lambda(2100)(J^P=\frac{7}{2}^-)$ states~\cite{Litchfield, Muller, Burgun, Tripp}, these results remain questionable due to low statistics,  and the results have not been corroborated. Therefore, the parameters of our simulation ($M(Y^*)$, $\Gamma(Y^*)$, and exponential $t$-slope values) were adjusted iteratively to match the data distributions. The final parameters for the $Y^*$ are $M=1.96$~GeV/{\it c}$^2$ and $\Gamma=220$~MeV/{\it c}$^2$. The $t$-slope values range from 1.11 to 2.64~(GeV/{\it c}$^2$)$^{-2}$ for the 11 photon energy bins from 2.75 to 3.85~GeV. After the simulation successfully  reproduced the data, the differential cross section results for the $\Xi^-$ were then extracted for the photon energy range of $2.75-3.85$~GeV. Due to limited statistics, only total cross sections in the  photon energy range of $3.85-4.75$~GeV have been extracted. 

Although the quark-diquark fusion mechanism was used to explain earlier $\Xi^-$ inclusive photoproduction data, hadronic degrees of freedom are of more relevance at energy range of this experiment. Partly due to the lack of data, there have been no theoretical predictions of the cascade production in exclusive photon-nucleon reactions until the production model developed by Nakayama {\it et~al.}~\cite{Oh} for the reaction $\gamma N \rightarrow K K \Xi$. Using an effective Lagrangian approach, the model incorporates various $t$, $u$, and $s$-channel processes, accounting for intermediate hyperon and nucleon resonances. The free parameters include the pseudoscalar-pseudovector (ps-pv) mixing parameter $\lambda$, the signs of the hadronic and electromagnetic transition coupling constants, the cutoff parameter $\Lambda_B$ and the exponent $n$ in the baryonic form factor $f_B$ [$f_B(p^2)=(\frac{n\Lambda_B^4}{n\Lambda_B^4+(p^2-m_B^2)^2})^n$, with $p$ denoting the baryon momentum and $m_B$ the baryon mass], and the product of the coupling constants $g_{N\Lambda K} g_{\Xi\Lambda K}$ for higher mass resonances. In their model, the ps-choice and pv-choice denote the extreme cases for the pseudoscalar-pseudovector (ps-pv) mixing parameter $\lambda$, {\it i.e.}, $\lambda=0$ for the pv-coupling choice and $\lambda=1$ for the ps-coupling choice.

\begin{figure}[htbp]
\vspace{80mm}
{\includegraphics{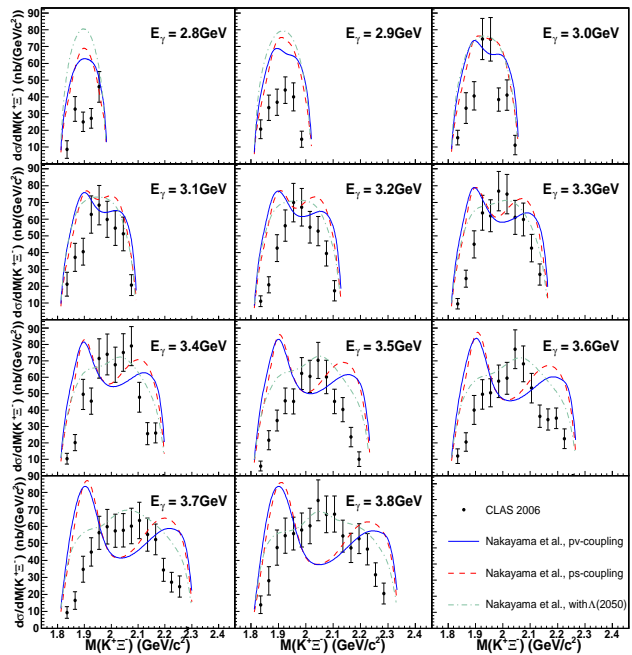}}
\vspace{5mm}
\caption{ (Color online) Differential cross section ($d\sigma/dM(K^+\Xi^-)$) results (including both statistical and systematic uncertainties) from the current work compared with model predictions from Ref.~\cite{Oh}. The solid curves correspond to the predictions with the pv-coupling, the dashed curves correspond to the ps-coupling choice, while the dot-dashed curves include an additional $\frac{3}{2}^+$ hyperon resonance at $2.05$~GeV/{\it c}$^2$ with $\Gamma=200$~MeV/{\it c}$^2$.}
\label{dsmkx}
\end{figure}

While Ref.~\cite{Oh} includes predictions using many variations of the parameters, the best agreement with our data requires $t$-channel processes involving at least one $J=\frac{3}{2}$ hyperon. Therefore, the more interesting differential cross sections would be $d\sigma/dM(K^+\Xi^-)$. Since there are two $K^+$ in the final state, both particles are included in the differential cross section extractions (Fig.~\ref{dsmkx}). The model of Ref.~\cite{Oh} includes the $\Lambda(1800)\frac{1}{2}^-$ and the $\Lambda(1890)\frac{3}{2}^+$, predicting a double humped behavior for the $M(\Xi^-K^+)$ spectra (Fig.~\ref{dsmkx}, solid and dashed curves). However, such a feature could potentially be smoothed out if an additional hypothetical hyperon state ($\Lambda(2050)\frac{3}{2}^+$, with $\Gamma=200$~MeV/{\it c}$^2$) is included in the model. The predictions agree with the data qualitatively when the additional $\Lambda(2050)$ state is included (Fig.~\ref{dsmkx}, dot-dashed curves). 

As for the hyperon states at lower masses, the data do not appear to support significant contributions from the  $\Lambda(1800)$ and the $\Lambda(1890)$, since the $K^+\Xi^-$ invariant mass spectra (Fig.~\ref{dsmkx}) peak significantly higher, at positions shifting according to the the photon energies. Whether these enhancements are due to hyperon states that decay to $K^+\Xi^-$, or simply larger phase space, could not be sufficiently determined by the current analysis. Further work by Mokeev {\it et al.} on the development of the JLAB-MSU phenomenological approach~\cite{MOKEEV} for exclusive reactions with three final state particles to incorporate the $K^+K^+\Xi^-$ channel is in progress, and may help to better determine the $\Xi^-$ photoproduction mechanism in the future.

\begin{figure}[htbp]
\vspace{80mm}
{\includegraphics{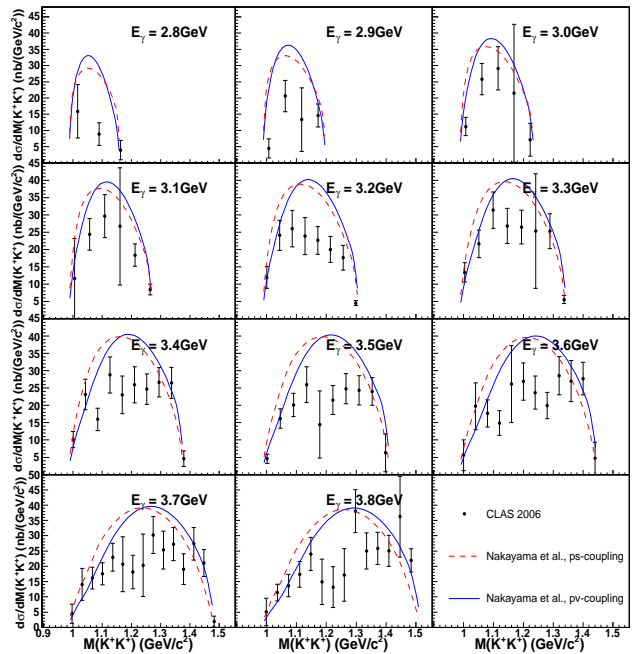}}
\vspace{5mm}
\caption{ (Color online) Differential cross section ($d\sigma/dM(K^+K^+)$) results (including both statistical and systematic uncertainties) from the current work compared with model predictions from Ref.~\cite{Oh}. The solid curves correspond to the predictions with the pv-coupling choice, while the dashed curves correspond to the ps-coupling choice.}
\label{dsmkk}
\end{figure}

Since no $S=+2$ meson system is believed to contribute to the reaction $\gamma p\rightarrow K^+K^+\Xi^-$, the $K^+K^+$ invariant mass spectrum is expected to be featureless, as is supported by both the data and the model of Nakayama {\t et al.}~\cite{Oh} (Fig.~\ref{dsmkk}).

\begin{figure}[htbp]
\vspace{80mm}
{\includegraphics{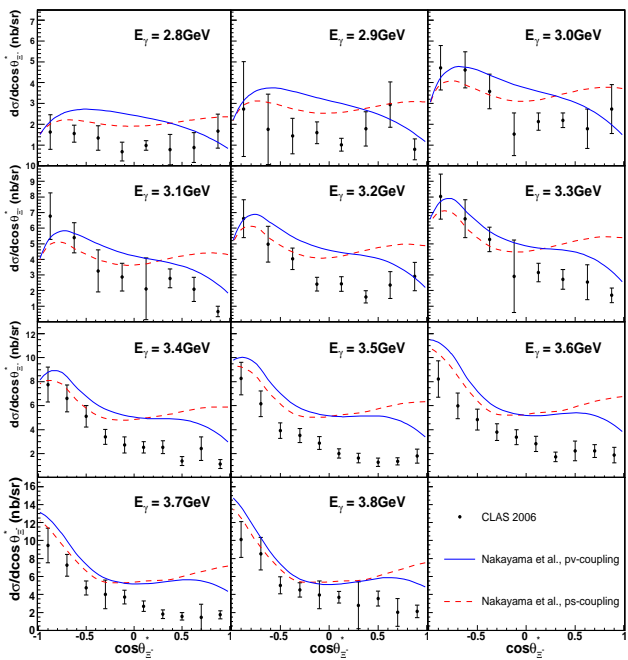}}
\vspace{5mm}
\caption[Differential cross section ($d\sigma/d\cos\theta^*_{\Xi^-}$)]{ (Color online) Differential cross section ($d\sigma/d\cos\theta^*_{\Xi^-}$) results (including both statistical and systematic uncertainties) from the current work compared with model predictions from Ref.~\cite{Oh}. The solid curves correspond to the predictions with the pv-coupling choice, while the dashed curves correspond to the ps-coupling choice.}
\label{dsdcos}
\end{figure}

\begin{figure}[htbp]
\vspace{80mm}
{\includegraphics{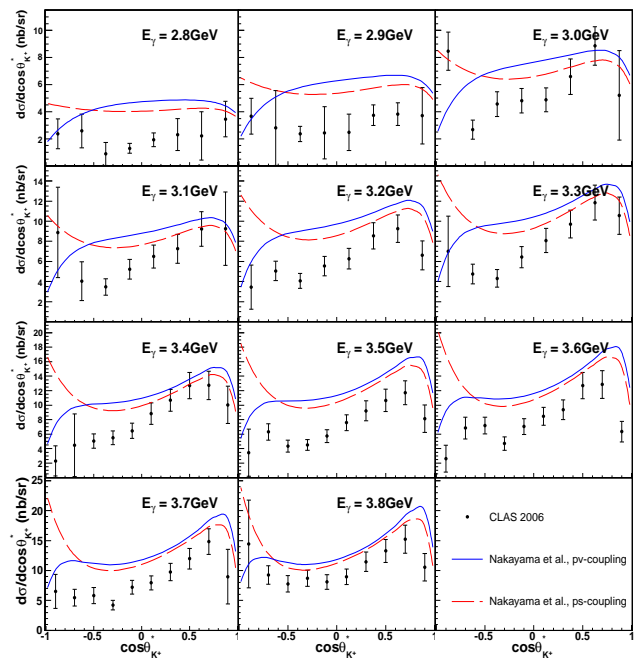}}
\vspace{5mm}
\caption{ (Color online) Differential cross section ($d\sigma/d\cos\theta^*_{K^+}$) results (including both statistical and systematic uncertainties) from the current work compared with model predictions from Ref.~\cite{Oh}. The solid curves correspond to the predictions with the pv-coupling choice, while the dashed curves correspond to the ps-coupling choice. }
\label{dsdcosk}
\end{figure}

The angular distributions of the $\Xi^-$ and $K^+$ in the photon-proton center-of-mass (c.m.) frame are also studied (Figs.~\ref{dsdcos},~\ref{dsdcosk}) . In Fig.~\ref{dsdcos}, the $\Xi^-$ angular distributions in the c.m. frame appear to be peaking backward for most of the energy bins, qualitatively agreeing with the predictions of Ref.~\cite{Oh}, which seems to overestimate the contributions from radiative transitional processes that tend to create forward-peaking features. As for the $K^+$ c.m. angular distributions (Fig.~\ref{dsdcosk}), the data exhibit a somewhat forward-peaking feature although it decreases in the most forward region. These angular distributions are consistent with the predictions that $\Xi$ photoproduction is dominated by $t$-channel hyperon processes.

The statistical uncertainties of the differential cross section results are around $15\%$. Systematic uncertainties due to the detector uncertainties, fiducial cuts, and flux normalization factors amount to around 10$\%$. Systematic uncertainties due to model dependence of the acceptance is extracted for each kinematic bin by comparing the values obtained using a range of simulation parameters. Such uncertainties are typically less than $5\%$, but may be as high as 10$\%$ for particular angular ranges such as the most forward or most backward regions of the detectors.

\begin{figure}[htbp]
\vspace{52mm}
{\includegraphics{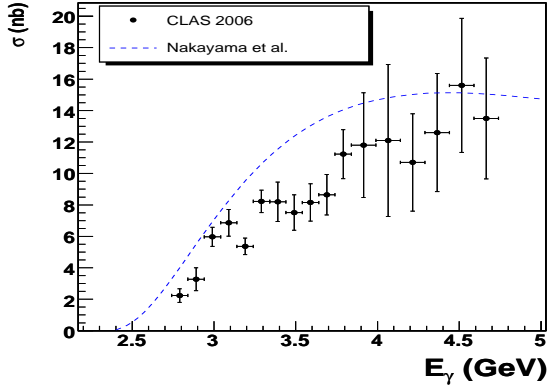}}
\vspace{2mm}
\caption{Total cross section of $\Xi^-$ results (including both statistical and systematic uncertainties) from the current work compared with model predictions from Ref.~\cite{Oh}.}
\label{total}
\end{figure}

After the differential cross sections for the $\Xi^-$ were obtained, the total cross sections (Fig.~\ref{total}) were determined as a function of $E_{\gamma}$ by integrating the differential cross sections. An additional systematic uncertainty, around 10$\%$, as a result of the integration is extracted by comparing the results of integrating the four different sets of differential cross sections. The $\Xi^-$ total cross section is determined to be around 2~nb at $E_{\gamma}=2.8$~GeV, and rises to about 11~nb at 3.8~GeV. The rising cross section with $E_{\gamma}$ is consistent with our conjecture for the simulation since higher photon energies simply provide more phase space, making it possible to produce other hyperon states that may decay to $K^+\Xi^-$. 

For $E_{\gamma}>3.85$~GeV, the statistics are limited and it is not feasible to fine-tune the simulation model to match the data in terms of various differential cross sections. Instead, the production of $\Xi^-$ is assumed to be of the same origin as that at $E_{\gamma}=3.8~$GeV. The total cross section results are then extracted in 6 energy bins for the $E_{\gamma}=3.85-4.75$~GeV region. Larger systematic uncertainties, estimated to be around $20\%$, are included for the total cross section results above $3.85$~GeV. Within uncertainties, the results are consistent with the continuation of the rise of $\sigma(E_{\gamma})$, slightly different from the flattening behavior predicted in Ref.~\cite{Oh}. However, it should be pointed out that Ref.~\cite{Oh} used earlier preliminary results reported in Ref.~\cite{Guo}, and it is likely the agreement between our data and the model could become significantly better.

It should be mentioned that the current results are higher than that reported earlier by CLAS ($3.5\pm1.1$~nb for $E_{\gamma}=3.0-3.9$~GeV, ~\cite{Price}), which were obtained from data with much lower statistics. The difference at the same energy range is $3.5\pm1.6$~nb, about 2 standard deviations from zero. This difference is mainly due to the different model for the CLAS acceptance and underestimated systematics of the previous measurement. 

\vspace{-1mm}
\section{$\Xi^-(1530)$ results}
\vspace{-1mm}

\begin{figure}[htbp]
\vspace{48mm}
{\includegraphics{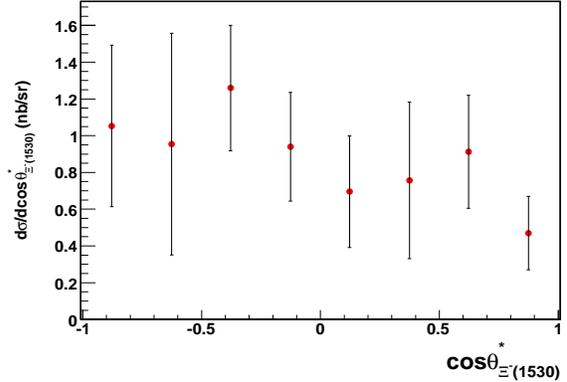}}
\vspace{5mm}
\caption[Differential cross sections for the $\Xi^-(1530)$]{Differential cross sections for the $\Xi^-(1530)$ in the photon energy range of 3.35-4.75~GeV. Both statistical and systematic uncertainties are included.}
\label{dsdcosx}
\end{figure}

The $700$ events in the $MM(K^+K^+)$ spectrum (Fig.~\ref{mmkk}) represent the highest statistics collected in exclusive photoproduction of the $\Xi(1530)$ to date. The $\Xi^-(1530)$ mass is found to be $1537.8\pm0.9\pm2.4$~MeV/{\it c}$^2$, while the width is $15.0\pm5.0$~MeV/{\it c}$^2$, both consistent with the previous measurements~\cite{YAO}. In the energy range of  $3.35-4.75~$GeV (there is no $\Xi^-(1530)$ signal below $3.35~$GeV, due to low acceptance and production rate), the $\Xi^-(1530)$ yields are extracted in eight cos$\theta^*_{\Xi^-(1530)}$ bins  in the c.m. frame to obtain the differential cross section, shown in Fig.~\ref{dsdcosx}. However, the statistics are not high enough to allow detailed model tuning for the simulation, which assumes a $t$-channel process that produces a hypothetical hyperon $Y^*$ ($M=2.155$~GeV/{\it c}$^2$, $\Gamma=160$~MeV/{\it c}$^2$, $t$-slope=$1.6$~(GeV/{\it c}$^{2}$)$^{-2}$) production that decays to $\Xi^-(1530)K^+$. The systematic uncertainty due to the model dependence of the CLAS acceptance is estimated to be around 20$\%$. The total cross section is then obtained by summing the differential cross section results and is $1.76\pm0.24\pm0.13$~nb for  $E_{\gamma}=3.35-4.75$~GeV, less than $20\%$ of that of the ground state in the comparable energy range. 

\begin{figure}[htbp]
\vspace{30mm}
{\includegraphics{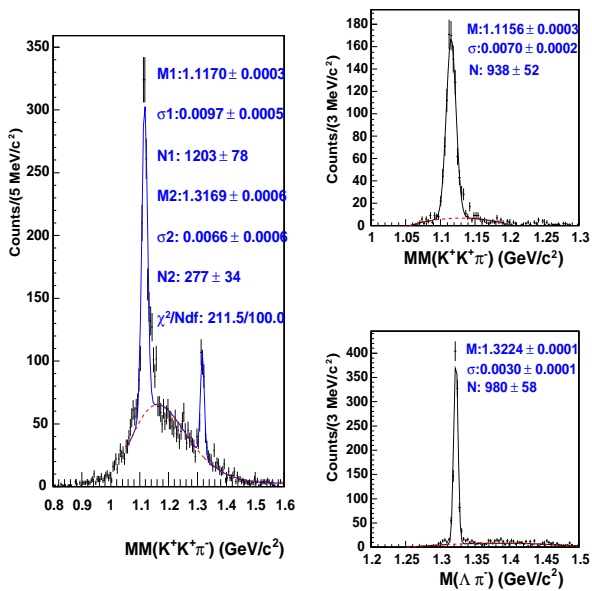}}
\vspace{50mm}
\caption{Left: ($K^+K^+\pi^-$) missing mass spectrum. The dashed background shape is obtained from events with an additional $\pi^+$ in the same event; Top right: ($K^+K^+\pi^-$) missing mass with a 3$\sigma$ cut on the $\Xi^-$ region (in the ($K^+K^+$) missing mass); Bottom right: ($\Lambda\pi^-$) invariant mass with a 3$\sigma$ cut on the $\Lambda$ region (in the ($K^+K^+\pi^-$) missing mass). Fitting parameter notation is the same as Fig.~\ref{mmkk}. 
}
\label{mmkkpi}
\end{figure}

To search for the excited cascade resonances, the reaction $\gamma p \rightarrow K^+ K^+ \pi^- [\Xi^0]$ has been studied. The main contributing background process is  $\Xi^-$ production because of the consequent decays  $\Xi^-\rightarrow \Lambda \pi^-$, and the missing particle from the $K^+K^+\pi^-$ system would be the $\Lambda$ (Fig.~\ref{mmkkpi}, top right).  It is interesting to note that the $\Xi^-$ signal reconstructed from the $\Lambda\pi^-$ invariant mass (Fig.~\ref{mmkkpi}, bottom right) has a much better resolution ($\sigma\sim$ 3~MeV/{\it c}$^2$) than using the missing mass technique (Fig.~\ref{mmkk}, $\sigma\sim$ 7~MeV/{\it c}$^2$). The $\Xi^-$ mass, as determined by the $\Lambda\pi^-$ invariant mass, is 1.3224~GeV/{\it c}$^2$, consistent with that identified from the reaction $\gamma p\rightarrow K^+K^+ (X)$ via missing mass. However, the statistics are much lower due to the low acceptance for the negative pion (around $10\%$). Therefore the $\Xi^-$ cross section results were extracted only using the $\gamma p\rightarrow K^+ K^+ (X)$ reaction. In addition, events with the $\pi^-$ coming from $\Lambda$ decay remain part of the background. To suppress this background, the vertex position from the $\pi^-$ is required to be within the target area because of the weak decay of the $\Lambda$. If an additional proton is detected and the $p\pi^-$ invariant mass falls close to the $\Lambda$ region, the event is removed from the final data sample. 

The $K^+K^+\pi^-$ events with an additional $\pi^+$ detected (about $20\%$ of the total $K^+K^+\pi^-$ events) are used to estimate the background, which is typically associated with those events where a $\pi^+$ or proton is misidentified as a $K^+$ (reactions such as $\gamma p\rightarrow K^+\Lambda(1520)$, $\Lambda(1520)\rightarrow \Lambda\pi\pi/\Sigma\pi$ can all contribute to this background). This empirical background peaks around $1.2$~GeV/{\it c}$^2$ in the $K^+K^+\pi^-$ missing mass spectrum, slightly overestimates the right shoulder of the $\Lambda$ peak, and in general describes the data well near the $\Xi^0$ peak(Fig.~\ref{mmkkpi}, left). The non-$\Xi^0$ event background is also explored by investigating those events orginating from outside of the target, which are less likely to be associated with the $\Xi^{-*}$ production. The results are qualitatively the same. 

\begin{figure}[htbp]
\vspace{40mm}
{\includegraphics{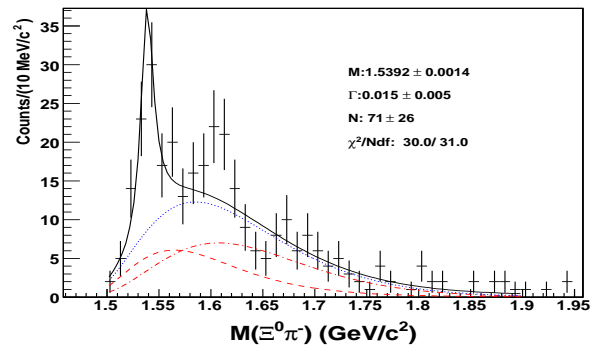}}
\vspace{5mm}
\caption{ 
(Color online) ($\Xi^0\pi^-$) invariant mass spectrum from events with $CL>0.1$. The dashed line is the non-$\Xi^0$ background obtained from events with $CL<0.1$, and the dash-dotted line is the $K^{*0}$ background defined by $\gamma p \rightarrow K^+ K^{*0} \Xi^0$ simulation. The dotted line is the total background as the sum of these two backgrounds. The $\Xi^-(1530)$ signal is parametrized by a p-wave Breit-Wigner function.}
\label{caspi}
\end{figure}

Finally, about 270 $\Xi^0$ events can be identified from the $K^+K^+\pi^-$ missing mass spectrum in addition to the dominant $\Lambda$ signal (Fig.~\ref{mmkkpi}, left). The $\Xi^0$ events are then kinematically fitted using the nominal $\Xi^0$ mass of $1.3148$~GeV/{\it c}$^2$. The final $M(\Xi^0\pi^-)$ spectrum is shown in Fig.~\ref{caspi}, where the $\Xi^-(1530)$ is visible. For those events that are associated with non-$\Xi^0$-production, events with low confidence level ($CL<10\%$) are used to study the background. The background obtained is included in the fit so that the total number of non-$\Xi^0$ events are within $10\%$ of the expected number of events. Using other methods to estimate this background as discussed earlier, and also side band events, yield similar results.

However, it should be pointed out that reactions such as $\gamma p \rightarrow K^+ K^{*0} \Xi^0, K^{*0}\rightarrow K^+ \pi^-$ and $\gamma p \rightarrow K^+ Y^*, Y^*\rightarrow Y^{*+}\pi^-\rightarrow K^+\pi^-\Xi^0$ may also contribute, complicating the interpretation of the spectrum. The knowledge of these processes is very limited, mostly due to the lack of data. The first process is simulated with a $t$-channel process of $K^*$ production with a heavy hyperon that decays to $K^+\Xi^0$, producing a background spectrum in the $\Xi^0\pi^-$ invariant mass as shown in the dot-dashed line of Fig.~\ref{caspi}. The spectrum was fitted with a p-wave Breit-Wigner function atop the non-$\Xi^0$-event background and the $K^{*0}$ background, yielding about $70~\Xi^-(1530)$ events (integrated from 1.50 to 1.8~GeV/{\it c}$^2$). The small enhancement around the 1.6~GeV/{\it c}$^2$ region has a significance of less than $2.5$ standard deviations, and will be further discussed in the next section.

The cross section of the $\Xi^-(1530)$ state can then be extracted and compared with the results obtained from the reaction $\gamma p \rightarrow K^+ K^+ (X)$ discussed earlier. As a consistency check, assuming the branching ratio of $\frac{BR(\Xi^{-*}\rightarrow \Xi^0\pi^-)}{BR(\Xi^{-*}\rightarrow (\Xi\pi)^-)}=\frac{2}{3}$, the $\Xi^{-}(1530)\rightarrow (\Xi\pi)^-$ cross section for the energy range of $3.35-4.75$~GeV has been determined to be $1.60\pm0.41\pm0.21$~nb for the $\Xi^-(1530)$, obtained from differential cross sections extracted in four angular bins of the $\Xi^0\pi^-$ system in the photon-proton c.m. frame. Within uncertainties, the branching ratio of the $\Xi\pi$ channel of the $\Xi^-(1530)$ decay extracted from this data, $0.91\pm0.30$, is consistent with the known value of $100\%$. 

\vspace{-5mm}
\section{$\Xi^0$ mass and the $\Xi$ doublet mass splitting}
\vspace{-3mm}
\vspace{0mm}
The mass of the $\Xi^0$, identified from the reaction $\gamma p\rightarrow K^+K^+\pi^- [\Xi^0]$, is measured to be $1316.9\pm 0.6 \pm 1.2$~MeV/{\it c}$^2$, higher the PDG value of $1314.83 \pm 0.2$~MeV/{\it c}$^2$~\cite{YAO}. The systematic uncertainty of 1.2~MeV/{\it c}$^2$ is derived from the dependence on the kinematic variables such as the $\Xi^0$ laboratory angles. The $\Xi$ doublet mass splitting can then be derived to be $5.4\pm 1.8$~MeV/{\it c}$^2$, consistent with the PDG value of $6.48\pm 0.24$~MeV/{\it c}$^2$. If the decay products of the $\Xi^0$ are detected, the mass can be determined from invariant mass instead of missing mass, and may lead to a better measurement of the $\Xi$ doublet mass splitting. However, it is impossible to achieve with the current statistics.

\vspace{-5mm}
\section{Discussions of $\Xi^*$}
\vspace{-3mm}
Among the lighter cascade resonances, the $\Xi(1620)$ is a controversial state that has only been reported in the $\Xi\pi$ channel, with very limited statistics; it is assigned only one star in the most recent PDG~\cite{YAO}. The reported mass, between 1600 to 1630~MeV/{\it c}$^2$, seems to be too low for the second excited cascade resonance according to the constituent quark model~\cite{Chao}. Earlier evidence~\cite{Briefel, Debellefon, Ross} has poor statistics. On the theoretical side, some dynamic models~\cite{Azimov, Ramos} have predicted a possible cascade resonance in the region of $1600$~MeV/{\it c}$^2$. In the framework of a unitary extension of chiral perturbation theory~\cite{Ramos}, the $\Xi(1620)$ emerged in the $\Xi\pi$ invariant mass with a width around 50~MeV/{\it c}$^2$, and is assigned to an octet together with the $N^*(1535)$, the $\Lambda(1670)$, and the $\Sigma(1620)$. These models clearly contradict the constituent quark model~\cite{Chao}. As for the $\Xi(1690)$, although it has recently been reported in the $\Xi\pi$ channel~\cite{Adamovich}, it has mostly been observed in the $\Lambda/\Sigma \; K^-$ decay, which has very low acceptance in the current experiment. 

In the two reactions reported here, there is no substantial signal for any excited cascade state beyond the $\Xi^-(1530)$. In the reaction  $\gamma p\rightarrow K^+K^+ (X)$, although the presence of the $\Xi^-(1530)$ is indubitable in the spectrum (Fig.~\ref{mmkk}), the data are consistent with background fluctuations in the $\Xi^-(1620)$ and the $\Xi^-(1690)$ regions. However, the absence of signals does not rule out the existence of these resonances, since it is likely that their production rate is too low to be observed due to the low photon energies and limited acceptance in our experiment. For the reaction $\gamma p\rightarrow K^+K^+ \pi^-[\Xi^0]$, the number of $\Xi^-(1530)$ events is consistent with the expectation when compared with the reaction  $\gamma p\rightarrow K^+K^+ [\Xi^-(1530)]$. In Fig.~\ref{caspi}, only the $\Xi^-(1530)$ signal is of statistical significance. In fact, the simulated $K^{*0}$ events also peak in the 1600~MeV/{\it c}$^2$ region, where the largest fluctuation occurs. Limited by the low statistics, the interference effect is challenging to quantify, making the interpretation of the data more difficult. It is also worth reminding the reader that processes such as the reaction  $\gamma p \rightarrow K^+ Y^*, Y^*\rightarrow Y^{*+}\pi^-\rightarrow K^+\pi^-\Xi^0$ are not included in the background simulation. To perform a full partial wave analysis and make more definite statements, an experiment with higher statistics is required.
\vspace{-5mm}
\section{Summary}
\vspace{-3mm}
The $\Xi$ doublet mass splitting is measured to be $5.4\pm 1.8$~MeV/{\it c}$^2$, consistent with the current global value of $6.48\pm 0.24$~MeV/{\it c}$^2$. In addition, the first detailed measurements of the $\Xi^-$ photoproduction cross sections have been obtained from the reaction  $\gamma p \rightarrow K^+ K^+ [\Xi^-]$. The $\Xi^-$ angular distributions and $K^+\Xi^-$ invariant mass spectra are consistent with a production mechanism of $Y^*\rightarrow \Xi^-K^+$ through a $t$-channel process. However, the current analysis is not sufficient to draw definite conclusions in terms of the production mechanism nor to determine the quantum numbers of the intermediate hyperon resonances.  The differential photoproduction cross sections of the  $\Xi^-(1530)$ have also been measured for the first time through the reaction  $\gamma p \rightarrow K^+ K^+ (X)$, and the $\Xi^-(1530)$ is also observed in the reaction  $\gamma p \rightarrow K^+ K^+ \pi^- [\Xi^0]$ as well. Although a small enhancement is observed in the $\Xi^0\pi^-$ invariant mass spectrum near the controversial 1-star $\Xi^-(1620)$ resonance,
it is not possible to determine its exact nature without a full partial wave analysis, due to the very limited statistics. This will be addressed by future, higher energy photon experiment using a hydrogen target that is currently planned in CLAS at Jefferson Laboratory~\cite{superg}.
\vspace{-5mm}
\section{Acknowledgment}
\vspace{-3mm}
We would like to acknowledge the outstanding efforts of the JLab Accelerator and the Physics Division staff, and especially the g11 running group, that made this experiment possible. We would like to thank B.~Nefkens and S.~Capstick for useful discussions.
This work was supported in part 
by 
the U.S. Department of Energy, 
the National Science Foundation, 
the Istituto Nazionale di Fisica Nucleare, 
the  French Centre National de la Recherche Scientifique, 
the French Commissariat \`{a} l'Energie Atomique, 
an Emmy Noether grant from the Deutsche Forschungs Gemeinschaft, the Research Corporation, and 
the Korean Science and Engineering Foundation.
Jefferson Science Associates (JSA) operates the 
Thomas Jefferson National Accelerator Facility for the United States 
Department of Energy under contract DE-AC05-060R23177. 

\end{document}